\documentclass[aps,prd,preprint,superscriptaddress,tightenlines,nofootinbib]{revtex4}

\usepackage{graphicx}
\usepackage{dcolumn}
\usepackage{bm}

\usepackage{latexsym}
\usepackage{amsmath}
\usepackage{xspace}
\newcommand{\MyLuminosity}{\ensuremath{281}\xspace}
\newcommand{\MyNDD}{\ensuremath{1.8}\xspace} 
\newcommand{\MyNDPDM}{\ensuremath{0.8}\xspace}  
\newcommand{\MyND}{\ensuremath{1.6}\xspace}  
\newcommand{\LNV}{LNV\xspace} 
\newcommand{\MyExtraColSep}{\hspace{0.5cm}}

\begin{document}


\preprint{CLNS 05/1928}       
\preprint{CLEO 05-16}         

\title{Search for Rare and Forbidden Decays
  $\bm{D^+ \rightarrow h^\pm e^\mp e^+}$}

\author{Q.~He}
\author{H.~Muramatsu}
\author{C.~S.~Park}
\author{E.~H.~Thorndike}
\affiliation{University of Rochester, Rochester, New York 14627}
\author{T.~E.~Coan}
\author{Y.~S.~Gao}
\author{F.~Liu}
\affiliation{Southern Methodist University, Dallas, Texas 75275}
\author{M.~Artuso}
\author{C.~Boulahouache}
\author{S.~Blusk}
\author{J.~Butt}
\author{O.~Dorjkhaidav}
\author{J.~Li}
\author{N.~Menaa}
\author{R.~Mountain}
\author{R.~Nandakumar}
\author{K.~Randrianarivony}
\author{R.~Redjimi}
\author{R.~Sia}
\author{T.~Skwarnicki}
\author{S.~Stone}
\author{J.~C.~Wang}
\author{K.~Zhang}
\affiliation{Syracuse University, Syracuse, New York 13244}
% ...
\collaboration{CLEO Collaboration} 
\noaffiliation


\date{August 10, 2005}

\begin{abstract} 
  Using $\MyNDPDM \times 10^{6}$
$D^+ D^-$ pairs collected with the CLEO-c detector
at the $\psi(3770)$ resonance, we have searched for
flavor-changing neutral current 
and lepton-number-violating 
decays of $D^+$ mesons to final states with dielectrons.
We find no indication of either, obtaining $90 \%$ confidence level
upper limits of
$\mathcal{B}(D^+ \rightarrow \pi^+ e^+ e^-)  <  7.4\times 10^{-6}$,
$\mathcal{B}(D^+ \rightarrow \pi^- e^+ e^+)  <  3.6\times 10^{-6}$,
$\mathcal{B}(D^+ \rightarrow K^+ e^+ e^-)  <  6.2\times 10^{-6}$,
and
$\mathcal{B}(D^+ \rightarrow K^- e^+ e^+)  <  4.5\times 10^{-6}$.
\end{abstract}


\maketitle

%
    Searches for rare-decay processes have played an important role in the
development of the Standard Model (SM).  The absence of flavor-changing 
neutral currents (FCNCs) in kaon decays led to the prediction of
the charm quark~\cite{goodetal+glr}, and the observation of $
B^0$--$\bar{B}^0$ mixing, a FCNC process, signaled the 
very large top-quark mass~\cite{Albrecht+rosner87}.  To date, rare and 
forbidden charm decays have been less informative and less extensively 
studied.  In this Letter we present searches for the FCNC 
decays~\cite{Burdman:2001tf,Fajfer:2001sa}
$D^+ \rightarrow \pi^+ e^+ e^-$ and
$D^+ \rightarrow K^+ e^+ e^-$, and the lepton-number-violating (\LNV) 
decays~\cite{Ali:2001gs}
$D^+ \rightarrow \pi^- e^+ e^+$ and $D^+ \rightarrow K^- e^+ e^+$.
(Charge-conjugate modes are implicit throughout this Letter.)
Short-distance FCNC processes in charm decays are much more highly
suppressed by the Glashow-Iliopoulos-Maiani
mechanism~\cite{gimktwz} than the corresponding 
down-type quark decays because of the range
of masses of the up-type quarks.  Observation of $D^+$ FCNC decays
could therefore provide indication of non-SM physics or of unexpectedly 
large rates for long-distance SM processes like 
$D^+ \rightarrow \pi^+ V$, $V \rightarrow e^+ e^-$,
with a real or virtual vector meson $V$.
The \LNV decays $D^+ \rightarrow \pi^- e^+ e^+$
and $D^+ \rightarrow K^- e^+ e^+$ are
forbidden in the SM.  They could be induced by a Majorana
neutrino, but with a branching fraction only of order $10^{-30}$.  Any
observation at experimentally accessible levels would be clear evidence 
of physics beyond the SM.
Past searches have set upper limits for the four dielectron decay modes in 
our study that are of order $10^{-4}$~\cite{Eidelman:2004wy}.  The limits 
for corresponding dimuon modes are about an order of magnitude more stringent.

%
%
The CLEO-c detector~\cite{Briere:2001rn,cleoiidetector,cleoiiidr,cleorich}
was used to collect a sample of \MyNDD million 
$e^+ e^- \rightarrow \psi(3770) \rightarrow D\bar{D}$ events
(\MyND million $D^\pm$ mesons) from an integrated luminosity of 
\MyLuminosity pb$^{-1}$ provided by the Cornell Electron Storage 
Ring (CESR).
From the interaction point out,
CLEO-c consists of a six-layer low-mass drift chamber,
a 47-layer central drift chamber,
a ring-imaging Cherenkov detector (RICH), and
a cesium iodide electromagnetic calorimeter,
all operating inside a $1.0$-T magnetic field provided by
a superconducting solenoidal magnet.  The detector provides 
acceptance of $93 \%$ of the full $4 \pi$ solid angle
for both charged particles and photons.
Charged particle identification (PID) is based on
information from the RICH detector,
the specific ionization ($dE/dx$) measured by the drift chamber,
and the ratio of electromagnetic shower energy to track momentum ($E/p$).
Background processes and the efficiency of signal-event selection
are estimated with  a GEANT-based~\cite{geant}
Monte Carlo (MC) simulation program.
Physics events are generated by EvtGen~\cite{evtgen}
and final-state radiation (FSR) is modeled by
the PHOTOS~\cite{photos} algorithm.
Signal events are generated with a phase-space model
as a first approximation of non-resonant FCNC and \LNV decays.

%
%
Candidate signal decays are reconstructed from well-measured charged-particle
tracks that are consistent in three dimensions with production at the 
$e^+e^-$ collision point.  Electrons with momenta of at least 200~MeV 
are identified with a likelihood ratio that combines $E/p$, $dE/dx$, and 
RICH information.  Charged kaons and pions with momenta of 50~MeV or
greater are selected based on $dE/dx$ and RICH information.
For each candidate decay of the form $D^+ \rightarrow  h^\pm e^\mp e^+$, 
where $h$ is either $\pi$ or $K$, we compute  the energy
difference $\Delta E =   E_\text{cand} - E_\text{beam}$ 
and the beam-constrained mass difference $\Delta M_\text{bc} =
\sqrt{E^2_\text{beam} - |\vec{p}_\text{cand}|^2} - M_{D^+}$,
where $E_\text{cand}$ and $\vec{p}_\text{cand}$ are the measured energy 
and momentum of the $h^\pm e^\mp e^+$ candidate, $E_\text{beam}$ is 
the beam energy, and $M_{D^+}$ is the nominal mass of the $D^+$ 
meson~\cite{Eidelman:2004wy}.  The resolution for these quantities 
is improved by recovering bremsstrahlung photons that are detected 
in the calorimeter within 100~mrad of electron trajectories.  This 
provides a signal-efficiency increase of 
$13\%-18\%$, depending on decay mode.
 
Events with $D^+$ candidates satisfying 
$-30~\text{MeV} \le \Delta M_\text{bc} < 30~\text{MeV}$
and $-100~\text{MeV} \le \Delta E < 100~\text{MeV}$
are selected for further study.  Within this region we define the 
``signal box'' to be 
$-5~\text{MeV} \le \Delta M_\text{bc} < 5~\text{MeV}$ and
$-20~\text{MeV} \le \Delta E < 20~\text{MeV}$, corresponding to 
$\pm 3 \sigma$ in each variable, as determined by MC simulation.
The remainder of the candidate sample was used to assess backgrounds.

The expected branching fraction for the long-distance 
decay $D^+ \rightarrow \pi^+ \phi \rightarrow \pi^+ e^+ e^-$ is 
within the sensitivity of this analysis ($\sim 10^{-6}$).  
We subdivide our candidates based on the mass squared of the 
final-state $e^+e^-$ (equal to the $q^2$ of the decay), with   
$0.9973~\text{GeV}^2 \le m^2_{e^+e^-} < 1.0813~\text{GeV}^2$ 
defining the $\phi$-resonant region.
We use this region both to veto the long-distance 
$D^+ \rightarrow \phi \pi^+ \rightarrow \pi^+ e^+ e^-$
contribution and to measure its branching fraction.

Backgrounds in the $D^+ \rightarrow  h^\pm e^\mp e^+$ candidate sample
arise from both $D \bar{D}$ and non-$D \bar{D}$ sources.  In $D \bar{D}$ 
events double semileptonic decays are dominant.  These typically
have four or fewer tracks (including two real electrons) and large
missing energy.  Potential peaking backgrounds from three-body hadronic
$D^+$ decays, such as $K^- \pi^+ \pi^+$, $\pi^+ \pi^+ \pi^-$, and 
$K^+ K^0_S$, are negligible because of the very small probability
of misidentifying charged hadrons as electrons in CLEO-c ($\sim 0.1\%$
per track), and because incorrect mass assignments result in $\Delta E$ 
outside the signal box.  In non-$D \bar{D}$ events, including continuum 
$e^+e^- \rightarrow q \bar{q}$ with $q \neq c$, $\tau$-pair events, and 
radiative return to $\psi(2S)$, non-peaking backgrounds arise from 
$\gamma$-conversion and Dalitz decays of $\pi^0$ and $\eta$.

We have performed a ``blind analysis.''
Signal-selection and background-suppression criteria were optimized
using MC simulation before we open the signal box
by minimizing the sensitivity variable
\begin{equation}
\mathcal{S} =
\frac{\sum_{n = 0}^{\infty} \mathcal{C}(n;N) \cdot \mathcal{P}(n;N)}
     {
       \epsilon
       \cdot
       N_{D^+}(\mathcal{L})
       },
\label{eq:sensitivity}
\end{equation}
where
$n$ is the observed number of events,
$N$ is the expected number of background events,
$\mathcal{C}$ is the 90\% confidence coefficient upper limit on the signal,
$\mathcal{P}$ is the Poisson probability,
$N_{D^+}$ is the number of charged $D$ mesons
(as a function of integrated luminosity $\mathcal{L}$),
and $\epsilon$ is the signal-selection efficiency.
The sensitivity variable $\mathcal{S}$ represents
the average upper limit on the
branching fraction that would be obtained from an ensemble of
experiments if the true mean for the signal were zero.
Sideband studies demonstrate that the MC simulation provides a 
good description of background events.

%
%
Background associated with $D$ semileptonic decays,
  mainly double semileptonic events with typically 4 or fewer tracks
  in the event with large missing energy
  (or semileptonic decay accompanied with $\gamma$-conversion
  or $\pi^0$ and $\eta$ Dalitz decays in the other side),
is suppressed by a requirement on the energy $E_\text{other}$,
the sum of the energies of all particles other than those making up
the signal candidate.
Small values of 
$E_\text{other}$ correspond to large values of missing energy in the 
event and are indicative of semileptonic decays in which neutrinos account
for significant undetected energy.  Optimization leads to different 
requirements on $E_\text{other}$ for different signal modes: 
$E_\text{other} > 1.0$ GeV for the $\pi^+ e^+ e^-$ final state,
$E_\text{other} > 1.3$ GeV for $K^+ e^+ e^-$,
and $E_\text{other} > 0.5$ GeV for the \LNV modes if the number of tracks
in the event is 4 or fewer.

%
%
Background events from $\gamma$-conversion and from $\pi^0$ and 
$\eta$ Dalitz decays are suppressed by rejecting $D^+$ candidates 
with low effective dielectron mass.
We use two kinds of dielectron effective mass squared
variables for this purpose:
$m^2_{e^+e^-}$ is computed for oppositely charged signal-side
electrons and $\acute{m}^2_{e^+e^-}$ is computed
for all combinations of one signal electron with any unused 
oppositely charged track.  We veto candidates if
$m^2_{e^+e^-} < 0.01~\text{GeV}^2$ 
or $\acute{m}^2_{e^+e^-} < 0.0025~\text{GeV}^2$. 

%
%
The decay mode $D^+ \rightarrow \pi^+ e^+ e^-$ is susceptible to 
background from $D^+ \rightarrow K^0_S e^+ \nu_e$ accompanied by a 
semileptonic decay of the other $D$.  This is suppressed by 
rejecting candidates when the signal $\pi^+$ and an oppositely charged 
track combine to give a mass $M_{\pi^+ \pi^-}$ that satisfies 
$-5~\text{MeV} \le M_{\pi^+\pi^-} - M_{K^0_S} < 5~\text{MeV}$,
where $M_{K^0_S}$ is the nominal $K^0_S$ mass~\cite{Eidelman:2004wy}.

After application of all background-suppression criteria,
our intention was to
eliminate multiple candidates
(candidates in excess of one per mode per charge per event) 
by selecting the smallest $|\Delta M_{\rm bc}|$ among
all that satisfy $-5~\text{MeV} \le \Delta M_{\rm bc} < 5~\text{MeV}$ 
and $-100~\text{MeV} \le \Delta E < 100~\text{MeV}$.
However, it turns out that there were \textit{no} multiple
candidate events.

The residual background and the efficiencies after application of all 
selection criteria have been determined by MC simulation and are given 
for the four signal modes in Table~\ref{table:table-result-sys}.
The model used to describe FCNC and \LNV decays is phase space.
The efficiency is observed to be quite uniform over the Dalitz plot, 
with the exception of the two corners at low $m^2_{ee}$, which
are depleted by the 200-MeV minimum-momentum requirement for electron
identification.

\begin{figure*}
\centering
\includegraphics[width=\textwidth]{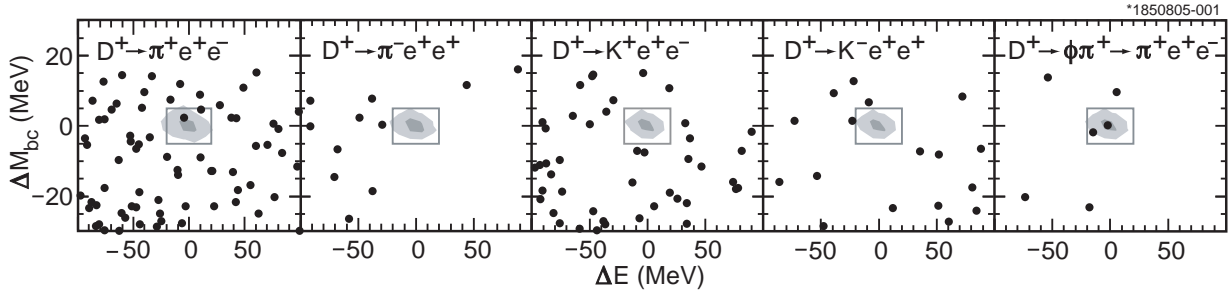}
\caption{\label{fig:final}Scatter plots of
$\Delta M_\text{bc}$ vs.\ $\Delta E$ obtained from data for each decay mode.
The signal region, defined
by $-20~\text{MeV} \le \Delta E < 20~\text{MeV}$ and 
$-5~\text{MeV} \le \Delta M_\text{bc} < 5~\text{MeV}$,
is shown as a box.  The two contours for each mode enclose regions 
determined with signal MC simulations
to contain $50\%$ and $85\%$ of signal events, respectively.
}
\end{figure*}

%
%
Scatter plots of $\Delta M_\text{bc}$ vs.\ $\Delta E$
for data events surviving all other cuts are shown in Fig.~\ref{fig:final}.
For $D^+ \rightarrow \pi^+ e^+ e^-$, two events lie in the signal box,
with an expected background of 1.99.
For all other FCNC or \LNV modes there are zero events in the
signal box.  With no evidence of a signal, we calculate
$90 \%$ confidence level (CL) upper limits (UL) on the branching fraction
for each mode from the observed number of events ($n$)
in the signal box, the signal-detection efficiency ($\epsilon$), and the
MC-estimated number of background events ($N$).  We follow
the Poisson procedure~\cite{Eidelman:2004wy} to calculate the 
$90\%$ confidence level coefficient ($\mathcal{C}(n;N)$)
upper limit on signal in the presence of expected background:
\begin{equation}
\text{UL}
  =
\frac{\mathcal{C}(n;N)}
     {\epsilon \cdot (2 \cdot \sigma_{D^+ D^-} \cdot \mathcal{L})} ,
\label{eq:ul}
\end{equation}
where $\sigma_{D^+ D^-}$~\cite{He:2005bs}
is the $e^+e^- \rightarrow \psi(3770) \rightarrow D^+ D^-$ cross section,
$\mathcal{L}$ is the integrated luminosity, and
$2 \cdot \sigma_{D^+ D^-} \cdot \mathcal{L}=$ 1.6 million is the number of 
charged $D$ mesons in our data sample.
Results are given in Table~\ref{table:table-result-sys}: we find no 
evidence of either FCNC or \LNV decays.
We separately measure the branching fraction for the resonant decay
$D^+ \rightarrow \pi^+ \phi \rightarrow \pi^+ e^+ e^-$,
finding two events in the signal region with an expected background of 0.04.

%
%
Systematic uncertainties in these results can be divided into two 
categories:  those related to background estimation and those arising 
from the signal-efficiency determination.

For the background uncertainty, only the $D^+ \rightarrow \pi^+ e^+ e^-$ 
mode needs to be considered, as the other modes have zero observed events 
and the uncertainty in the expected number of background events does not 
affect their upper limits.
%
%
For $D^+ \rightarrow \pi^+ e^+ e^-$,
we compared the background estimate from the MC simulations
with that of data from the 
$\Delta E$-$\Delta M_\text{bc}$ sideband.  The sideband estimate of the
background in the signal box is about
one standard deviation ($\sigma$)
higher than the MC
estimate.  Therefore our upper limit based on the MC background estimate
is conservative; the upper limit with the sideband-estimated background
would be 5\% lower.

Sources of uncertainties that are common to all results are
the number of $D^+$ ($-3.2\%$, $+4.5\%$),
tracking ($\pm 1 \%$ per track or $\pm 3\%$ total),
PID ($\pm 2.3\%$), 
and FSR ($\pm 4.0 \%$ for $\pi^\pm e^\mp e^+$, 
$\pm 3.3 \%$ for $K^+ e^+ e^-$, $\pm 3.5 \%$ for $K^- e^+ e^+$,
and $\pm 4.4 \%$ for $\pi^+ \phi \rightarrow \pi^+ e^+ e^-$,
estimated by comparing the efficiency
before and after bremsstrahlung recovery).

Uncertainties in signal efficiency due to background-suppression cuts are 
estimated by comparing the efficiency
before and after the cuts are applied:
$\pm 5.2 \%$ ($\pi^+ e^+ e^-$),
$\pm 1.1 \%$ ($\pi^- e^+ e^+$),
$\pm 7.3 \%$ ($K^+ e^+ e^-$),
$\pm 1.0 \%$ ($K^- e^+ e^+$),
and
$\pm 0.9 \%$ ($\pi^+ \phi \rightarrow \pi^+ e^+ e^-$).

Uncertainty from using the phase-space model
(as a first approximation for non-resonant decays)
for the FCNC and \LNV signal efficiency estimation is
assessed by (somewhat arbitrarily)
taking one quarter of the fraction of phase space
which has non-uniform efficiency due to the electron identification
momentum cut-off (200~MeV):
$\pm 2.8 \%$ ($\pi^\pm e^\mp e^-$) and $\pm 3.8 \%$ ($K^\pm e^\mp e^-$).

For the results in Table~\ref{table:table-result-sys}, we increase 
the upper limits to account for systematic uncertainties 
by decreasing the efficiency by 
$1 \sigma_\text{syst}$ (combined systematic uncertainty).
\begin{table}
\centering
\caption{\label{table:table-result-sys}Efficiencies ($\epsilon$),
background estimates ($N$),
observed yields ($n$),
combined systematic uncertainties ($\sigma_\text{syst}$),
and branching fraction results for four FCNC and \LNV decay modes
and for the resonant decay 
$D^+ \rightarrow \pi^+ \phi \rightarrow \pi^+ e^+ e^-$.
Branching-fraction UL values are all at 90\% CL.
}
\begin{tabular}{c@{\MyExtraColSep}r@{\MyExtraColSep}r@{\MyExtraColSep}r@{\MyExtraColSep}r@{\MyExtraColSep}r}
\hline
\hline
Mode& $\epsilon$ (\%)& $ N $& $n$& $\sigma_\text{syst}$ (\%)& $\mathcal{B}$ ($10^{-6}$)\\ 
\hline\hline
$\pi^+ e^+ e^-$&
$36.41$&
$1.99$&
$2$&
$8.7$&
$<7.4$
\\\hline
$\pi^- e^+ e^+$&
$43.85$&
$0.48$&
$0$&
$7.1$&
$<3.6$
\\\hline
$K^+ e^+ e^-$&
$26.18$&
$1.47$&
$0$&
$10.0$&
$<6.2$
\\\hline
$K^- e^+ e^+$&
$35.44$&
$0.50$&
$0$&
$7.2$&
$<4.5$
\\\hline\hline
$\pi^+ \phi (e^+ e^-)$&
$46.22$&
$0.04$&
$2$&
$7.4$&
$2.7^{+3.6}_{-1.8} \pm 0.2$
\\\hline
\hline
\end{tabular}
\end{table}

%
%
In summary, we find no evidence for non-Standard-Model physics.
There is no evidence either for the two rare (FCNC) decays
or for the two forbidden (LNV) decays of charged $D$ mesons to
three-body final states with dielectrons.
Finding no evidence for signals, we set $90 \%$
confidence level upper limits:
\[
\begin{array}{lcr}
\mathcal{B}(D^+ \rightarrow \pi^+ e^+ e^-) & < & 7.4\times 10^{-6}\\
\mathcal{B}(D^+ \rightarrow \pi^- e^+ e^+) & < & 3.6\times 10^{-6}\\
\mathcal{B}(D^+ \rightarrow K^+ e^+ e^-) & < & 6.2\times 10^{-6}\\
\mathcal{B}(D^+ \rightarrow K^- e^+ e^+) & < & 4.5\times 10^{-6}
\end{array}
\]
Our results for these dielectron modes are significantly more restrictive 
than previous limits, and reflect sensitivity comparable to 
the searches for dimuon modes ~\cite{Eidelman:2004wy}.
Due to the dominance of long-distance effects in FCNC modes, we separately
measure the branching fraction of the resonant decay
$D^+ \rightarrow \pi^+ \phi \rightarrow \pi^+ e^+ e^-$,
obtaining
$\mathcal{B}(D^+ \rightarrow \phi \pi^+\rightarrow \pi^+ e^+ e^-) = (2.7^{+3.6}_{-1.8} \pm 0.2) \times 10^{-6}$.
This is consistent with the product of known world
average~\cite{Eidelman:2004wy} branching fractions,
$
\mathcal{B}(D^+ \rightarrow \phi \pi^+ \rightarrow \pi^+ e^+ e^-)
=
      \mathcal{B}(D^+ \rightarrow \phi \pi^+)
      \times
      \mathcal{B}(\phi \rightarrow e^+ e^-)
      = [(6.2\pm 0.6) \times 10^{-3}] \times [(2.98\pm 0.04) \times 10^{-4}]
      = (1.9 \pm 0.2) \times 10^{-6}
$.


We gratefully acknowledge the effort of the CESR staff 
in providing us with excellent luminosity and running conditions.
This work was supported by the National Science Foundation
and the U.S. Department of Energy.

\end{document}